\documentclass[aps,prd,twocolumn,groupedaddress,nofootinbib]{revtex4-2}

\usepackage{amssymb}
\usepackage{amsmath}
\usepackage{graphicx}
\usepackage{enumerate}
\usepackage{mathtools}
\usepackage{color}
\usepackage{bbold}
\usepackage{subfigure}
\usepackage{slashed}
\usepackage{xcolor}
\usepackage{bm}
\usepackage{multirow}
\usepackage{pifont}
\usepackage{aas_macros}
\usepackage{comment}

\begin{document}

\title{Collisional flavor swap with neutrino self-interactions}

\author{Chinami Kato}
 \affiliation{Faculty of Science and Technology, Tokyo University of Science, 2641 Yamazaki, Noda-shi, Chiba 278-8510, Japan}
 \email{ckato@rs.tus.ac.jp}
\author{Hiroki Nagakura}%
\affiliation{%
 National Astronomical Observatory of Japan, 2-21-1 Osawa, Mitaka, Tokyo 181-8588, Japan}%
\author{Lucas Johns}%
\affiliation{%
 Departments of Astronomy and Physics, University of California, Berkeley, CA 94720, USA}%

\date{\today}

\begin{abstract}
Neutrinos play pivotal roles in determining fluid dynamics, nucleosynthesis, and their observables in core-collapse supernova (CCSN) and binary neutron star merger (BNSM). In this paper, we present a novel phenomenon, collisional flavor swap, in which neutrino-matter interactions trigger the complete interchange of neutrino spectra between two different flavors, aided by neutrino self-interactions. 
We find that a necessary condition to trigger the collisional swap is occurrences of resonance-like collisional flavor instability.
In cases where neutrino self-interactions substantially dominate over the collision rate, the collisional swap occurs in the entire neutrino energy spectrum, while intriguing energy dependent features can emerge after the completion of flavor swap.
Since flavor swaps correspond to the most extreme case in flavor conversions, they have a great potential to affect CCSN and BNSM phenomena.
\end{abstract}

\maketitle

\section{Introduction}
Exploring neutrino flavor conversions driven by neutrino-self interactions (collective neutrino oscillations) \cite{pantaleone1992,samuel1993,sigl1993,sigl1995} is a key frontier in the study of core-collapse supernovae (CCSNe) and binary neutron star mergers (BNSMs).
The detailed investigation has been motivated by theoretical indications that flavor conversions ubiquitously occur in CCSN- \cite{dasgupta2017,tamborra2017,nagakura2019,milad2019,milad2020,morinaga2020,abbar2020,abbar2020b,capozzi2020,nagakura2021d,harada2022,akaho2022,Xiong2022,nagakura2023} and BNSM \cite{wu2017,wu2017b,george2020,Li2021,sherwood2022,grohs2022,nagakura2023b} environments.
Their physical properties, however, remain shrouded in a mystery, despite their growing attention.

Some recent studies have also suggested that collisional flavor instability (CFI), a new type of collective neutrino oscillations, can occur in optically thick regions of CCSNe \cite{Xiong2022,Liu2023b}
and BNSMs \cite{Xiong2022b}.
Our understanding of CFI has been matured rapidly based on linear stability analysis  (see, e.g., \cite{Liu2023,Xiong2022b}), but much less work has been done on their non-linear properties. We also note that the previous studies have ignored diagonal components of collision terms \cite{johns2021,johns2022,Lin2022}, which potentially discards some important characteristics of CFI. In fact, we shall show that the diagonal components play a pivotal role on characterizing non-linear dynamics of CFI.

In this paper, we present a novel phenomenon in nonlinear phases of neutrino flavor conversions, named as collisional flavor swap (or collisional swap). 
Here we use "swap" to refer to the simultaneous interchange of different flavors, i.e., more extreme than the flavor equipartition. 
There are two noticeable properties in collisional swap: (1) the timescale is much faster than neutrino-matter interactions, since the growth of flavor conversions in the early phase is associated with the resonance-like CFI \cite{Xiong2022b,Liu2023}:
(2) the collisional swap can occur in isotropic neutrino distributions in momentum space, indicating that the interplay with fast neutrino-flavor conversion is not necessary \cite{johns2022}. 

We stress that the collisional swap is distinct from other swap phenomena in the literature such as spectral swap \cite{duan2006,duan2007} and MSW effects. Occurrences of rapid and vigorous flavor conversions by CFI in regions where neutrinos and matter are strongly coupled can affect all relevant physics in these phenomena, including fluid dynamics, nucleosynthesis, and observable signals such as gravitational waves and neutrinos \cite{Xiong2020,Just2022,fernandez2022,Ehring2023,Ehring2023b,fujimoto2023}. This exhibits the possibility of a large impact of collisional swap on both theories and observations for CCSNe and BNSMs.

\section{Dynamical simulations}
We start with presenting results of dynamical simulations for collisional swap by solving quantum kinetic equations (QKEs) of neutrinos. We solve the energy-dependent QKEs under isotropic and spatial homogeneous neutrino background,
\begin{eqnarray}
i\frac{\partial}{\partial t} \rho(t,E_\nu) = [H_{\nu\nu},\rho(t,E_\nu)]+iC, \label{rho_eq}
\end{eqnarray}
with the density matrix $\rho$, the neutrino energy $E_\nu$, the collision term $C$, and the Hamiltonian potential,
\begin{eqnarray}
H_{\nu\nu} = \sqrt{2}G_F\int dV_q\left[ \rho(t,E_\nu) - \bar{\rho}^\ast(t,E_\nu) \right]. \label{H}
\end{eqnarray}
$G_F$ and $V_q$ are the Fermi constant and the volume element in momentum space, respectively. 
In this study, we neglect vacuum and matter potentials just for simplicity.
We also assume the two-flavor system consisting of electron neutrinos ($\nu_e$) and heavy-leptonic neutrinos ($\nu_x$), in which case $\rho$ is a 2$\times$2 matrix, i.e.,
\begin{eqnarray}
\rho = \begin{pmatrix}
\rho_{ee} & \rho_{ex} \\
\rho_{xe} & \rho_{xx}
\end{pmatrix}.
\end{eqnarray}
For antineutrinos, we use "-" expression and replace $\rho\rightarrow\bar{\rho}$, $C\rightarrow\bar{C}$ and $H_{\nu\nu}\rightarrow \bar{H}_{\nu\nu} =-H_{\nu\nu}^\ast$ in Eqs.~\ref{rho_eq} and \ref{H}.

In the collision term, the electron- and positron captures by free protons and neutrons, respectively, are included as emission processes, while absorption processes are also taken into account by their inverse reactions. $\nu_x$ reactions are not included in this study. Reaction rates of these processes are computed by following \cite{bruenn1985} under given a fluid distribution. Baryon mass density, temperature, and electron fraction are set as $10^{12}\ {\rm g/cm^3}$, $6.4$ MeV and $0.1$, respectively (see Tab.~\ref{tab:condition}), while the similar matter state has been observed in recent CCSN models \cite{furusawa2023}.
We also note that these parameters are chosen so that the $\nu_e$ chemical potential becomes zero, which corresponds to a necessary condition for occurrences of a resonance-like evolution in CFI (unless number densities of
$\nu_x$ and their antipartner ($\bar{\nu}_x$) are largely different from each other, whose cases are not considered in this paper, though). We employ a nuclear equation-of-state \cite{furusawa2017} to obtain all necessary thermodynamical quantities for computing weak reaction rates.

As an initial condition, we assume that $\nu_e$ and $\bar{\nu}_e$
are in thermal and chemical equilibrium with matter, whereas $\nu_x$ and $\bar{\nu}_x$ is assumed to be Fermi-Dirac distributions with the chemical potential of -2~MeV.
The choice of the chemical potential of $\nu_x$ is based on our CCSN model \cite{nagakura2019b}.
We consider the region outside the energy sphere, at which neutrino emission and absorption are balanced each other\footnote{ We follow the convection in \cite{Janka2017}, which distinguishes energy- and transport spheres for $\nu_x$ (and $\bar{\nu}_x$). Unlike $\nu_e$ and $\bar{\nu}_e$, there is a scattering dominant region. In such a region, they are not in thermal- and chemical equilibrium with matter, but their angular distributions are nearly isotropic due to scatterings with nucleons. The transport (or scattering) sphere is located outside the energy one, and it is defined as the sphere where $\nu_x$ transits to free streaming.}.
In this region, $\nu_x$ and $\bar{\nu}_x$ undergo large numbers of scatterings mainly by nucleons, which leads to the negative chemical potential \cite{kato2020}.
It should be noted, however, that $\nu_x$ radiation field is sensitive to neutrino-matter interactions and multi-dimensional effects such as proto-neutron star convection \cite{nagakura2020}, which would affect occurrences of CFI \cite{Liu2023b,akaho2024}. It is, hence, necessary to inspect $\nu_x$ when we assess occurrences of collisional swap in CCSN models.
We add very small perturbations in off-diagonal components of density matrix ($10^{-6}$ compared to electron-type neutrinos) to trigger flavor conversions.

\begin{table*}
    \centering
    \small
    \begin{tabular}{cccc}
        \hline\hline
        density & temperature & electron fraction & chemical potential \\ \hline
        $10^{12}~{\rm gcm^{-3}}$ & 6.4~MeV & 0.1 & 0~MeV ($\nu_e$, $\bar{\nu}_e$), -2~MeV ($\nu_x$, $\bar{\nu}_x$) \\ \hline\hline
    \end{tabular}
    \caption{Setups in our model.}
    \label{tab:condition}
\end{table*}

We assess the stability of neutrino distributions by following the prescription in \cite{Liu2023}, and confirmed that CFI occurs with the growth rate of $5\times10^{-3}~{\rm cm^{-1}}$. The associated timescale of CFI ($t_{\rm CFI}$) is $\sim 10^{-5}$ shorter than the time scale of neutrino-matter interaction ($t_{\rm col}$). This exhibits that the flavor instability corresponds to a resonance-like CFI, whose time scale can be roughly estimated as $t_{\rm CFI}\sim\sqrt{G_F n_{\nu}\gamma}$, where $n_{\nu}$ and $\gamma$ denote the number density of neutrinos and energy-averaged reaction rates of neutrino-matter interactions, respectively. We solve the QKEs by using MC code \cite{kato2021}, in which we employ a uniform energy grid from $0$ to $100$~MeV with 100 grid points. We also carry out the same simulation but with twice the energy resolution (200 grids). We find that the result is almost identical to that with our standard resolution (the error of neutrino number density is less than $0.1 \%$). Hereafter, we, hence, discuss the collisional swap based on the model with the standard resolution.

\begin{figure}
\begin{flushleft}
    \includegraphics[width=8.5cm,clip]{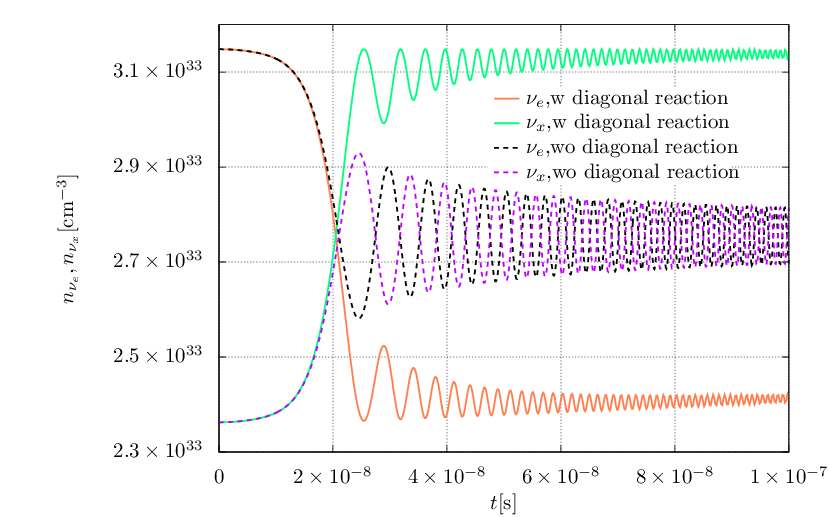}
    \caption{Time evolution of number densities of neutrinos. Red- and green-solid lines denote $\nu_e$ and $\nu_x$ ($n_{\nu_e}$ and $n_{\nu_x}$), respectively.
    Solid and dotted lines denote cases with and without diagonal components of collision term, respectively; see text for more details.
    }
    \label{number_dens}
\end{flushleft}
\end{figure}

Solid lines in Fig.~\ref{number_dens} draw the dynamics of collisional swap, while we omit to show those in antineutrinos, since their evolution is almost identical to neutrinos. $\nu_e$ and $\nu_x$ substantially shuffle at $t\sim2\times10^{-8}$~s, and then the flavor swap almost completes by $t\sim1\times10^{-7}$~s.

Before discussing the physical process of collisional swap in detail, we make an interesting comparison to the case with no collision term in diagonal elements; the results are shown as dotted lines in Fig.~\ref{number_dens}.
In the early phase, the time evolution of flavor conversions is almost identical to the case with diagonal collision terms, which is consistent with linear analysis. However, they deviate each other from $t\sim2\times10^{-8}$~s, corresponding to the time when the number of neutrinos of two flavors become nearly equal. In the case without diagonal collision terms, the system converges to a flavor equipartition with oscillations. This exhibits that the diagonal elements in collision terms are key elements to understand collisional swap, which can also be shown analytically (see below).

\begin{figure*}
    \centering
    \includegraphics[width=\textwidth]{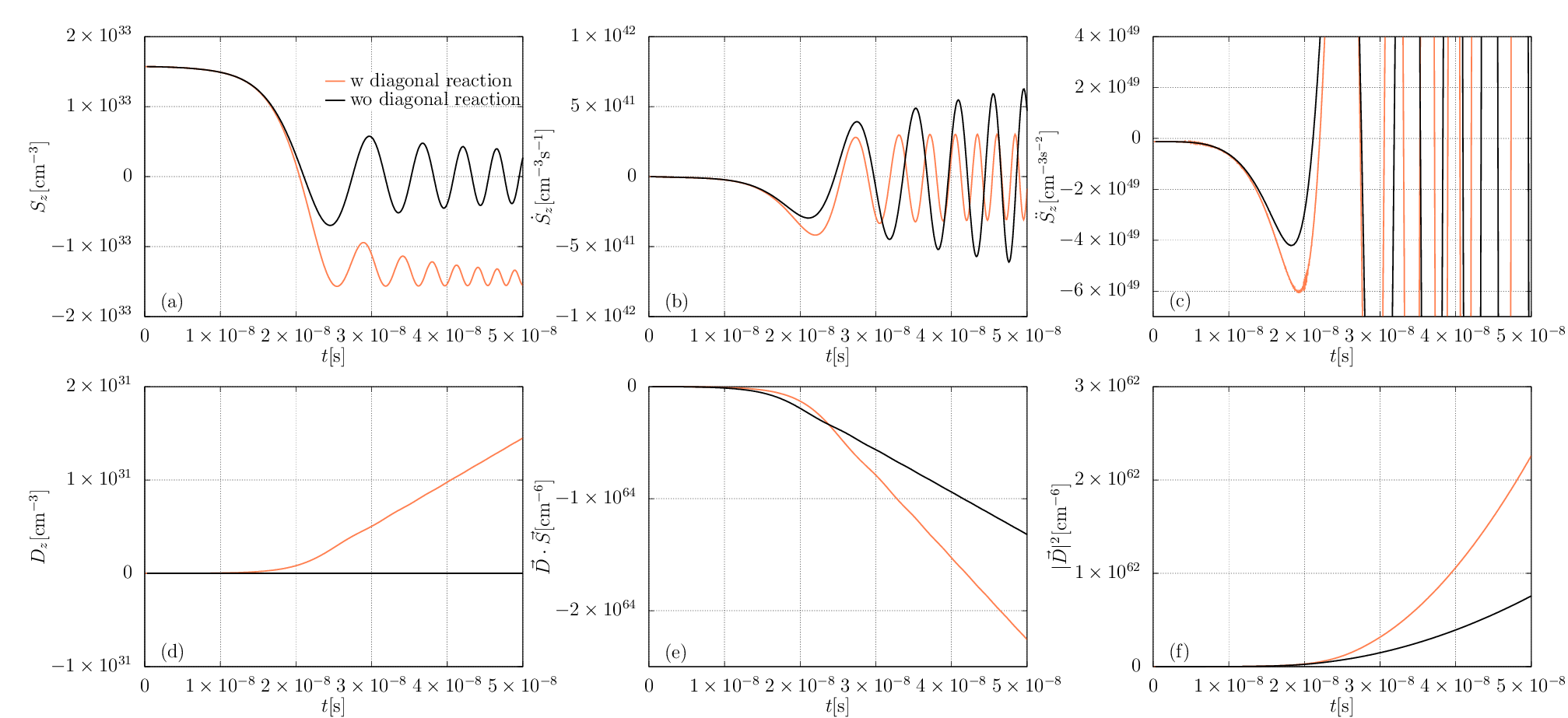}
    \caption{Time evolution of (a) $S_z$, (b) $\dot{S}_z$, (c) $\ddot{S}_z$, (d) $D_z$, (e) ${\bm D}\cdot{\bm S}$ and (f) $|{\bm D}|^2$. Red and black lines denote cases with and without diagonal components of collision term, respectively.}
    \label{fig:Szddot}
\end{figure*}

\section{Analytic arguments}\label{sec:anaArgu}
We discuss the collisional swap in terms of polarization vectors in flavor spaces, which are defined by $\rho \equiv P_0I/2+\bm{P}\cdot\bm{\sigma}/2$ with
\begin{eqnarray}
\bm{P} &=& (2{\rm Re}\rho_{ex},-2{\rm Im}\rho_{ex},\rho_{ee}-\rho_{xx}), \\
\bm{\bar{P}} &=& (2{\rm Re}\bar{\rho}_{ex}, 2{\rm Im}\bar{\rho}_{ex},\bar{\rho}_{ee}-\bar{\rho}_{xx}),
\end{eqnarray}
$P_0=\rho_{ee}+\rho_{xx}$, the unit matrix $I$ and the Pauli-matrix vector $\bm{\sigma}$.
It should be mentioned that $\bm{P}$ and $\bm{\bar{P}}$ depend on neutrino energy.
In this expression, the QKEs are
\begin{eqnarray}
  \frac{\partial}{\partial t}\bm{P}= \bm{H}_{\nu\nu}\times \bm{P} + \Gamma\left(\bm{P}_{eq}-\bm{P}\right) + \Gamma\left(P_{0,eq}-P_0\right)\bm{z}, \\
  \frac{\partial}{\partial t}\bm{\bar{P}}= \bm{H}_{\nu\nu}\times \bm{\bar{P}} + \bar{\Gamma}\left(\bm{\bar{P}}_{eq}-\bm{\bar{P}}\right) + \bar{\Gamma}\left(\bar{P}_{0,eq}-\bar{P}_0\right)\bm{z}, 
\end{eqnarray}
with $\Gamma=\Gamma_e/2$ and $\bar{\Gamma}=\bar{\Gamma}_e/2$; $\Gamma_e$ ($\bar{\Gamma}_e$) denotes the reaction rate for $\nu_e$ (electron antineutrinos, $\bar{\nu}_e$), while we consider the situation with $\Gamma > \bar{\Gamma}$ due to neutron rich environment; $\bm{z}$ is the unit vector of $z$-axis in flavor space; the index of "eq" indicates the quantities in the thermal equilibrium.
The Hamiltonian vector is
\begin{eqnarray}
\bm{H_{\nu\nu}} = \mu \int dV_q \left(\bm{P}-\bm{\bar{P}}\right),
\end{eqnarray}
with $\mu=\sqrt{2}G_F$.

To capture the essential features of collisional swap, we consider an energy-integrated form of QKE, while the detailed discussion about energy dependence is deferred to Sec.~\ref{sec:enedepe}. The QKEs can be approximated as
\begin{eqnarray}
 \frac{\partial}{\partial t}\bm{P}_{\rm int} &\sim& -\mu \bm{\bar{P}}_{\rm int}\times\bm{P}_{\rm int}
  + \Gamma_{\rm ave} \left(\bm{P}_{{\rm int},eq}-\bm{P}_{\rm int}\right) \nonumber \\
  &+& \Gamma_{\rm ave} \left(P_{0 {\rm int},eq}-P_{0 {\rm int}}\right)\bm{z}, \\
  \frac{\partial}{\partial t}\bm{\bar{P}}_{\rm int} &\sim& -\mu \bm{\bar{P}}_{\rm int}\times\bm{P}_{\rm int}
  + \bar{\Gamma}_{\rm ave}\left(\bm{\bar{P}}_{{\rm int},eq}-\bm{\bar{P}}_{\rm int}\right) \nonumber \\
  &+& \bar{\Gamma}_{\rm ave}\left(\bar{P}_{0{\rm int},eq}-\bar{P}_{0{\rm int}}\right)\bm{z},
\end{eqnarray}
where
\begin{eqnarray}
\bm{P}_{\rm int} \equiv \int dV_q \bm{P}, \label{eq:eneintP}
\end{eqnarray}
and $\Gamma_{\rm ave}$ denotes the collision rate at the average energy of neutrinos.
Throughout this section, we omit to show these subscripts.

For convenience, we discuss the collisional swap based on $\bm{S} \equiv \bm{P}+\bar{\bm{P}}$, and $\bm{D} \equiv \bm{P}-\bar{\bm{P}}$ instead of $\bm{P}$ and $\bm{\bar{P}}$.
The QKEs can be rewritten in terms of $\bm{S}$ and $\bm{D}$ as
\begin{eqnarray}
\dot{\bm{S}} &\sim& \mu\bm{D}\times\bm{S} \nonumber \\
&& + \frac{\Gamma+\bar{\Gamma}}{2}\left(\bm{S}_{eq}-\bm{S}+\left(S_{0,eq}-S_0\right)\bm{z}\right) \nonumber \\
&& + \frac{\Gamma-\bar{\Gamma}}{2}\left(\bm{D}_{eq}-\bm{D}+\left(D_{0,eq}-D_0\right)\bm{z}\right), \label{Sdot} \\
\dot{\bm{D}} &\sim& \frac{\Gamma-\bar{\Gamma}}{2}\left(\bm{S}_{eq}-\bm{S}+\left(S_{0,eq}-S_0\right)\bm{z}\right) \nonumber \\
&& + \frac{\Gamma+\bar{\Gamma}}{2}\left(\bm{D}_{eq}-\bm{D}+\left(D_{0,eq}-D_0\right)\bm{z}\right). \label{Ddot} 
\end{eqnarray}
In the initial condition, $\bm{S}$ is headed in the positive direction along $z$-axis (but slightly tilted from the $z$-axis due to perturbations), while $\bm{D}$ is embedded in $x-y$ plane (i.e., $D_z = 0$), and its $x-$ and $y$ components represent initial perturbations.

Here we consider reasonable approximations in Eqs.~\ref{Sdot}~and~\ref{Ddot} so as to make the problem analytically tractable. We assume that neutrino self-interactions are much stronger than neutrino-matter interactions, which guarantees $ t_{\rm CFI} \ll t_{\rm col}$. Since the collisional swap occurs in nonlinear phases of CFI, its dynamical timescale is also $t_{\rm CFI}$. This indicates that, given our initial conditions, $S_0 \sim S_{0,eq}$, $D_0 \sim D_{0,eq}$, and $|\bm{S}| \gg |\bm{D}|$ are reasonable approximations during the collisional swap. By using these conditions, we can approximate Eqs.~\ref{Sdot}~and~\ref{Ddot} as
\begin{eqnarray}
\dot{\bm{S}} &\approx& \mu \bm{D}\times\bm{S}, \label{Sdot_simple}\\
\dot{\bm{D}} &\approx& \frac{\Gamma-\bar{\Gamma}}{2} \left(\bm{S}_{eq}-\bm{S}\right). \label{Ddot_simple}
\end{eqnarray}
From Eqs.~\ref{Sdot_simple} and \ref{Ddot_simple}, we obtain the following relations,
\begin{eqnarray}
|\bm{S}| &\approx& |\bm{S}_{eq}|, \label{Sconserv} \\
\dot{D}_z &\approx& \frac{\Gamma-\bar{\Gamma}}{2}\left(|\bm{S}|-S_z \right), \label{Dzevo} \\
\partial_t\left(\bm{D}\cdot\bm{S}\right) &\approx& \frac{\Gamma-\bar{\Gamma}}{2}\left(\bm{S}\cdot\bm{S}_{eq}-|\bm{S}|^2\right), \label{DdotS} \\
\dot{|\bm{D}|^2} &\approx& ( \Gamma-\bar{\Gamma}) ( D_z |\bm{S}| - \bm{D}\cdot\bm{S} ), \label{D2eq} \\
\ddot{S}_z &\approx& \mu^2\left(\bm{D}\cdot\bm{S
}\right)D_z - \mu^2|\bm{D}|^2S_z, \label{Szevo}
\end{eqnarray}
which highlight essential features of collisional swap. It should be stressed that these relations hold in nonlinear phases (but $t \ll t_{\rm col}$).

Eq.~\ref{Sdot_simple} guarantees that the norm of $\bm{S}$ is constant in time; hence it can be given by the initial condition (Eq.~\ref{Sconserv}). On the other hand, $D_z$ monotonically increases with time, which can be derived from Eq.~\ref{Dzevo} with conditions of $\Gamma > \bar{\Gamma}$ and $|\bm{S}| > S_z$. This also indicates that $D_z$ becomes positive at $t>0$ (see also Fig.~\ref{fig:Szddot}(d)). The trend is also intuitively understandable as follows. Once flavor conversions happen, both $\nu_e$ and $\bar{\nu}_e$ are reduced. The restoring force of $\nu_e$ to the equilibrium state (i.e., $\nu_e$ emission) is stronger than $\bar{\nu}_e$ due to $\Gamma > \bar{\Gamma}$, accounting for the increase of $D_z$. Using similar arguments, $\bm{D}\cdot\bm{S}$ decreases with time, implying that it becomes negative at $t>0$ (see also Fig.~\ref{fig:Szddot}(e)). We can also derive that $|\bm{D}|^2$ monotonically increases with time from Eq.~\ref{D2eq} and the above relations (see also Fig.~\ref{fig:Szddot}(f)).

The time evolution of $\bm{S}$, in particular for the $z$-component, exhibits the dynamics of collisional swap in Fig.~\ref{fig:Szddot}(a). Here, we focus on its second derivative (see Eq.~\ref{Szevo}). The first term in the right hand side of Eq.~\ref{Szevo} is negative, because of $\bm{D}\cdot\bm{S}<0$ and $D_z > 0$ at $t>0$. As a result, $S_z$ separates from the $z$-axis initially, i.e., facilitating flavor conversions. The second term also accelerates the flavor conversion (since it is negative) until $S_z=0$, implying that $\bm{S}$ falls to the $x-y$ plane without undergoing decelerations. When $\bm{S}$ reaches the $x-y$ plane, it still moves towards the negative direction of $z$-axis (see Fig.~\ref{fig:Szddot}(b) at $t\lesssim2.5\times10^{-8}$~s). After $S_z<0$, however, the second term in the right hand side of Eq.~\ref{Szevo} flips the sign, i.e., decelerating flavor conversions. On the other hand, the first term remains to be negative, implying that $S_z$ is persistently pushed towards $-|\bm{S}|$ (i.e., flavor swap). The competition between the first and second terms causes $S_z$ oscillation (see also Fig.~\ref{fig:Szddot}(c)), but the persistent force by the first term makes the system settle into $S_z/|\bm{S}| \sim -1$, leading to the flavor swap.

As described above, the first term in the right hand side of Eq.~\ref{Szevo} is a key player to achieve the collisional swap. Here, we show that neglecting diagonal components of collision term results in qualitatively different outcome. This is attributed to the fact that Eq.~\ref{Dzevo} in the case is rewritten as

\begin{eqnarray}
\dot{D}_z &\sim& 0. \label{eq:Dz_nodiag}
\end{eqnarray}
This indicates that $D_z(=0)$ is constant in time, and consequently the first term in the right hand side of Eq.~\ref{Szevo} (which is not changed in the case without diagonal components of collision term) remains to be zero at $t \ge 0$ (see also Fig.~\ref{fig:Szddot}(d)). This also implies that $\bm{D}$ is always embedded in $x-y$ plane, making $S_z$ oscillates between positive and negative, since $\bm{S}$ rotates around $\bm{D}$ (see Eq.~\ref{Sdot_simple}).

A few remarks are in order. The first term in the right hand side of Eq.~\ref{Szevo} can be negative only if $D_z \ll S_z$. This is attributed to the fact that $\bm{D} \cdot \bm{S}$ and $D_z$ have the same sign at $t=0$, but either of them needs to flip the sign during CFI. This is only possible for cases with $D_z \ll S_z$, i.e., the resonance-like CFI.

Second, the above discussion based on $\bm{S}$ and $\bm{D}$ can be applied to cases with $\nu_x\neq\bar{\nu}_x$. From the similar argument, the collisional swap can occur, if the condition of reasonance-like CFI is satisfied. This is an important indication to CCSN- and BNSM study, since $\nu_x$ and $\bar{\nu}_x$ spectra are, in general, different from each other due to high-order corrections of weak interactions (e.g., effects of weak-magnetism \cite{Horowitz:2001xf}). The deviation would be more prominent if on-shell muons appear in these environments \cite{Bollig:2017lki,fischer2020}.

\begin{figure}
    \centering
    \includegraphics[width=\columnwidth]{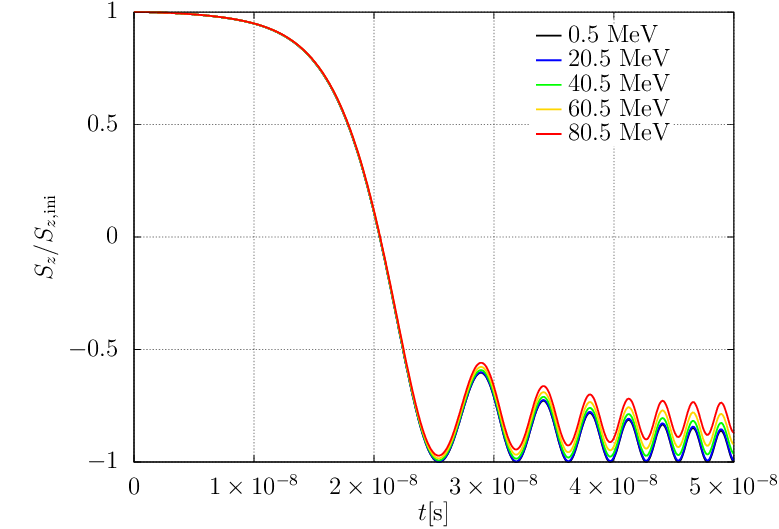}
    \caption{Time evolution of $S_z$ in each neutrino energy. Colors distinguish neutrino energy. The vertical scale is normalized by the initial value.}
    \label{Sz_ene}
\end{figure}

\section{Energy dependence}\label{sec:enedepe}
In this section, we discuss the energy dependence of collisional swap. Let us first show the time evolution of $S_z$ for some selected neutrino energies in Fig.~\ref{Sz_ene}. As shown in this figure, the collisional swap occurs for all energies of neutrinos, and their time evolution is nearly identical. One might think that this result is counterintuitive because the collision rate, which corresponds to a driving force of collisional swap, depends on energy. Below, we explain why there are less energy-dependent features in collisional swap.

One thing we do notice here is that the energy-dependent collision rate affects the dispersion relation for the CFI, but the resultant growth rate of flavor instability is identical among all neutrinos (see, e.g., \cite{Liu2023}), that guarantees the energy independence of CFI in the very early phase. However, this explanation is not sufficient for collisional swap, since it occurs beyond the linear phase. We hence consider the energy-dependence directly from the nonlinear QKEs as below.

Similar as Sec.~\ref{sec:anaArgu}, we assume a condition that the neutrino self-interaction is much stronger than neutrino-matter interaction, and we focus on the phase of $t \ll t_{\rm col}$. These conditions allow us to approximate the time evolution of $\bm{S}$ (similar to Eq.~\ref{Sdot_simple}) as,
\begin{eqnarray}
\dot{\bm{s}}(E_\nu) \sim \mu\bm{D}_{\rm int}\times\bm{s}(E_\nu).
\label{eq:enedepeS}
\end{eqnarray}
with $\bm{s} \equiv \bm{S}/S_{{\rm ini},z}$. In this expression, we do not omit the subscript of "${\rm int}$" and we emphasize that $\bm{s}$ is defined as energy-dependent quantity by explicitly showing neutrino energy, $E_\nu$. As can be seen in Eq.~\ref{eq:enedepeS}, the time evolution of $\bm{s}$ is driven by the energy-integrated quantity, $\bm{D}_{\rm int}$. We also note that all neutrino energies have the identical initial condition, $\bm{s}=(\xi_x,\xi_y,1)$, where $\xi_x$ and $\xi_y$ represent the initial small perturbations in $x$- and $y$ components, respectively. This argument guarantees that the time evolution of $\bm{s}$ are identical among all energies. We can also obtain the time evolution of $s_z$ as,
\begin{eqnarray}
|\bm{s}| &\approx& 1, \label{Sconserv_enedepe} \\
\ddot{s}_z &\approx& \mu^2\left(\bm{D}_{\rm int}\cdot\bm{s
}\right)D_{{\rm int},z} - \mu^2|\bm{D}_{\rm int}|^2s_z, \label{Szevo_enedepe}
\end{eqnarray}
which are essentially the same equations as Eqs.~\ref{Sconserv}~and~\ref{Szevo}. We can apply the same argument as described in Sec.~\ref{sec:anaArgu}, and consequently $s_z$ can reach $-1$, that exhibits the energy-independent collisional swap. One of the important points along this discussion is that $\bm{D}(E_\nu)$ does not directly contribute the collisional swap, whereas its energy-integrated quantity, $\bm{D}_{\rm int}$, is responsible for it. This argument also illustrates that the energy-dependence of $\Gamma$ and $\bar{\Gamma}$ does not directly affect the collisional swap, and their energy-averaged one affects the swap through the time evolution of $\bm{D}_{\rm int}$ (see Eq.~\ref{Ddot_simple}).

\begin{figure}
    \centering
    \includegraphics[width=\columnwidth]{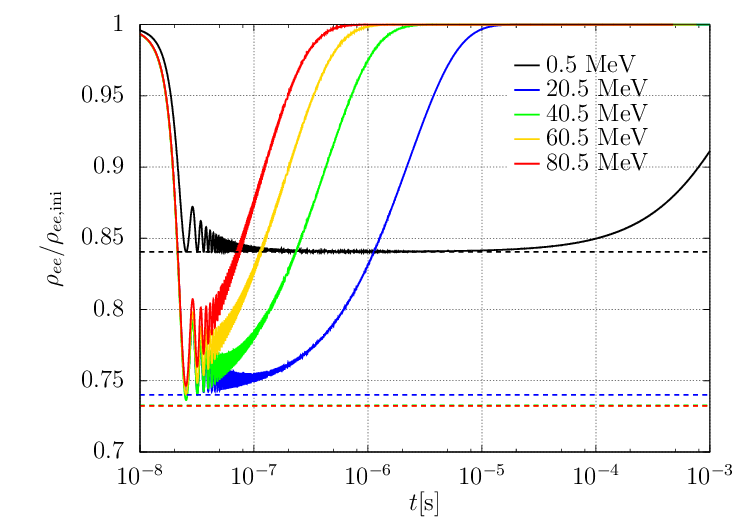}
    \caption{
    Long-term evolution of $\nu_e$ in each neutrino energy.
    Colors represent neutrino energy. The vertical scale is normalized by the initial value. 
    To guide our eyes, we also show the initial value of $\nu_x$ with dashed lines, and the color code is the same as $\nu_e$.
    }
    \label{late_Evo}
\end{figure}

The energy dependence of flavor evolution appears after the collisional swap is completed. After completing the swap, flavor conversions subside because the flavor state is stable with respect to the CFI. On the other hand, it can not be an asymptotic state because  charged-current reactions make $\nu_e$ and $\bar{\nu}_e$ restore the thermal equilibriums at $t \gtrsim t_{\rm col}$. This trend is clearly displayed in Fig.~\ref{late_Evo}, in which the time evolution of $\nu_e$ for some selected neutrino energies are portrayed. At the initial phase, $\nu_e$ undergoes a collisional swap regardless of neutrino energies, and then they return to the initial position at $t \gtrsim t_{\rm col}$. Such a late time evolution of neutrinos naturally generates the energy-dependent features, since the restoring speed hinges on the reaction rate.

This argument suggests an intriguing possibility, spectral-swap, may arise in real CCSN- and BNSM environments. One thing we do notice here is that the low-energy $\nu_e$ and $\bar{\nu}_e$ would not reach their equilibrium states in realistic system, because the advection time scale is shorter than the collision one. As a result, $\nu_e$ and $\bar{\nu}_e$ energy spectra in the low-energy region can not be restored and they would remain to be depleted, which may form a spectral-swap-like structure.

It should be noted, however, that there remain some uncertainties for this spectral swap phenomenon associated with collisional swap.
For instance, the canonical spectral swap involves a discrete change of energy spectrum (see, e.g., \cite{duan2006,duan2007}), but the survival probability of neutrinos should change continuously in the energy spectrum for cases with collisional swap\footnote{We note that the canonical spectral swap would also be smoothed by multi-angle effects as demonstrated in \cite{Wu2015}}.
We also note that there are other elements (e.g., other neutrino-matter interactions) that make the transition smooth. Although this phenomenon is worthy of further investigation, the consistent treatment of neutrino advection, multiple channels of neutrino-matter interactions, and flavor conversions is necessary to address this issue. We leave the detailed study for a future work.

\section{Conclusions}
We present that neutrino-matter interactions can lead to neutrino flavor swap between two flavors, aided by neutrino self-interactions. Different from the linear phase, diagonal components of collision term play a key role in the dynamics. The necessary condition for the collisional swap is a resonance-like CFI, which is realized when both $\nu_e$ and $\bar{\nu}_e$ number densities are nearly equal to each other (under the assumption of $\nu_x = \bar{\nu}_x $). We also find that the collisional swap occurs regardless of neutrino energies, as long as neutrino self-interactions are much stronger than the collision rate at each energy.

The collisional swap corresponds to the most extreme case in flavor conversions, and it would change the equilibrium state among neutrinos and fluids. 
We also note that the large flavor conversions in optically thick regions facilitate neutrino cooling \cite{nagakura2023b}, that
would give substantial impact on CCSN- and BNSM dynamics. 
In fact, resonance-like CFIs have been observed in recent simulations, see, e.g., \cite{Xiong2022b}.
The present study also shows that the phenomenon similar to the spectral swap could emerge in the late phase (after the collisional swap is completed), although more work is needed to quantify how much the survival probability is changed in the energy spectrum and how much it can affect the astrophysical consequences. These important issues will be addressed in our forthcoming papers.

\begin{acknowledgements}
C.K. is supported by JSPS KAKENHI Grant Numbers JP20K14457 and JP22H04577.
H.N. is supported by Grant-inAid for Scientific Research (23K03468) and also by the NINS International Research Exchange Support Program. We also acknowledge support from a HPCI System Research Project (Project ID: 230033).
L.J. is supported by NASA Hubble Fellowship grant number HST-HF2-51461.001-A awarded by the Space Telescope Science Institute, which is operated by the Association of Universities for Research in Astronomy, Incorporated, under NASA contract NAS5-26555.
\end{acknowledgements}

\bibliography{collisional_swap}

\end{document}